\documentclass[twocolumn,aps,prb,showpacs,amsmath,superscriptaddress]{revtex4}
\usepackage{amssymb}
\usepackage{mathrsfs}
\usepackage{graphicx}
\usepackage{float}
\usepackage{txfonts}
\begin{document}
\title{Topological invariants for phase transition points of one-dimensional $\mathbb{Z}_2$ topological systems}
\author{Linhu Li}
\affiliation{Beijing Computational Science Research Center, Beijing, 100089, China}
\affiliation{CeFEMA, Instituto Superior T$\acute{e}$cnico, Universidade de Lisboa, Av. Rovisco Pais, 1049-001 Lisboa, Portugal}
\author{Chao Yang}
\affiliation{Beijing National Laboratory for Condensed Matter
Physics, Institute of Physics, Chinese Academy of Sciences, Beijing
100190, China}
\author{Shu Chen}
\affiliation{Beijing National Laboratory for Condensed Matter
Physics, Institute of Physics, Chinese Academy of Sciences, Beijing
100190, China}
\affiliation{Collaborative Innovation Center of Quantum Matter, Beijing, China}
\begin{abstract}
  We study topological properties of phase transition points of two topologically non-trivial $\mathbb{Z}_2$  classes (D and DIII) in one dimension by assigning a Berry phase defined on closed circles around the gap closing points in the parameter space of momentum and a transition driving parameter. While the topological property of the $\mathbb{Z}_2$ system is generally characterized by a $\mathbb{Z}_2$ topological invariant, we identify that it has a correspondence to the quantized Berry phase protected by the particle-hole symmetry, and then give a proper definition of Berry phase to the phase transition point. By applying our scheme to some specific models of class D and DIII, we demonstrate that the topological phase transition can be well characterized by the Berry phase of the transition point, which reflects the change of Berry phases of topologically different phases across the phase transition point.
\end{abstract}
\maketitle
\date{today}
\section{Introduction}

The investigation on topological phases and phase transitions, especially the ones of topological insulators and superconductors, is one of the most attractive topics in theoretical study of condensed matter physics in the past decade \cite{review1,review2,bookTI,reviewTSC1,reviewTSC2}. While conventional quantum phases are described by continuous order parameters, topological ones are characterized by quantized topological invariants, which correspond to topological properties of occupied quantum states. These topological properties are robust under small adiabatic deformations of the Hamiltonian, and changing them requires so called topological phase transitions, which do not accompany symmetry breaking in contrast to conventional quantum phase transitions \cite{QPT}, but are symbolized by gap closing at some specific points in the Brillouin zone (BZ). However, the topological invariants are ill-defined at topological phase transition points, as they are usually defined for each quantum state protected by nonzero energy gaps.

According to the Altland-Zirnbauer (AZ) symmetry classification \cite{AZ}, a 1D fermion system can be classified into ten different symmetry classes, and five of them support topological phases \cite{classification}. In our recent works \cite{TransitionPointPRB,TransitionPointEPL}, it is shown that a topological phase transition in the $\mathbb{Z}$-type system can be well characterized by the topological property of the phase transition point by introducing a topological invariant defined on closed curve in the extended parameter space surrounding it.
Particularly, for one-dimensional (1D) $\mathbb{Z}$-type topological systems, it is shown that the winding number of the transition point defined in the parameter space of momentum and transition driving parameter can reveal the topological difference of phases on two sides of the transition point. While the topological property of phase transition point for the $\mathbb{Z}$-type system is already well understood, it is still not clear how to characterize the topological phase transition in $\mathbb{Z}_2$-type systems through a proper definition of topological invariant for the transition point, which is the initial research motivation of this work. It is known that topological phases of the two 1D topologically nontrivial BdG classes (D and DIII) are characterized by $\mathbb{Z}_2$ topological invariants,
which are equivalent to quantized Berry phases \cite{Berry} protected by the particle-hole symmetry (PHS) of systems \cite{Hatsugai,Budich1,Budich2}. This fact leads to the speculation that phase transitions in these classes may also have a connection to Berry phases of transition points.

As one of the simplest 1D topologically nontrivial models, the Kitaev's p-wave superconductor model \cite{Kitaev} has been widely studied since its discovery, and is an ideal example to illustrate the topological properties of 1D systems \cite{Budich1,Budich2,ZDWang,Sen,Lang,Cai,Sacramento,Kitaev,DeGottardi}. In this paper, we examine the symmetry-protected quantized Berry phase of 1D BdG classes, and define the Berry phase for the phase transition points in the 2D parameter space of momentum and the transition driving parameter. This Berry phase is defined on a closed path around the gapless point, which enables us to avoid the ill-definition problem at the transition point. We then demonstrate that the defined Berry phase is also quantized due to an extended symmetry in the parameter
space similar to the PHS, and has a correspondence to the $\mathbb{Z}_2$ topological phase transition. Finally, we apply our scheme to
study two extended versions of Kitaev's p-wave superconductor model, which belong to the two $\mathbb{Z}_2$ topological classes in 1D respectively.

\section{Topological invariant of one-dimensional BdG classes}
The Hamiltonian of the BdG class fulfills the particle-hole symmetry $\mathcal{C} H \mathcal{C}^{-1}=-H$, where $\mathcal{C}$ is an anti-unitary operator. For the simplest two-band BdG system, the Hamiltonian can be written as
\begin{eqnarray}
H=\psi_k^{\dagger}h(k)\psi_k,
\end{eqnarray}
where $\psi_k^{\dagger}=(\hat{c}_k,\hat{c}_{-k}^{\dagger}$) and $
h(k)=h_x\sigma_x+h_y\sigma_y+h_z\sigma_z$ with $\sigma$ the Pauli matrices acting on the vector $\psi_k$. The particle-hole symmetry is represented by
\begin{eqnarray}
\sigma_x \mathcal{K} h(k) \mathcal{K} \sigma_x=-h(-k),\label{PHS}
\end{eqnarray}
with $\mathcal{K}$ the complex conjugation operator\cite{classification}. The topological properties of this model can be described by the Majorana number\cite{Kitaev}. Defining the Majorana operator in momentum space
\begin{eqnarray}
\hat{a}_k=\hat{c}_k+\hat{c}_{-k}^{\dagger},~\hat{b}_k=(\hat{c}_k-\hat{c}_{-k}^{\dagger})/i
\end{eqnarray}
with $\hat{a}_k^{\dagger}=\hat{a}_{-k}$ and $\hat{b}_k^{\dagger}=\hat{b}_{-k}$,  we have
\begin{eqnarray}
h(k)\propto \left(
\begin{array}{cc}
\hat{a}_k &\hat{b}_k
\end{array}\right)
B(k)\left(
\begin{array}{c}
\hat{a}_{-k}\\
\hat{b}_{-k}
\end{array}\right),
\end{eqnarray}
and
\begin{eqnarray}
B(k)=\left(
\begin{array}{cc}
h_x & h_y-ih_z\\
h_y+ih_z & -h_x
\end{array}\right).
\end{eqnarray}
The Majorana number $M$ is then determined by the Pfaffian of $B(k)$ at $k=0$ and $\pi$:
\begin{eqnarray}
M&=&\mathrm{sgn}[\mathrm{Pf}B(\pi)]\mathrm{sgn}[\mathrm{Pf}B(0)]\nonumber\\
&=&\mathrm{sgn}(h_z(\pi))\mathrm{sgn}(h_z(0)),
\end{eqnarray}
where $M=1$ and $-1$ correspond to topologically trivial and non-trivial cases respectively.

\begin{figure}
\includegraphics[width=0.8\linewidth]{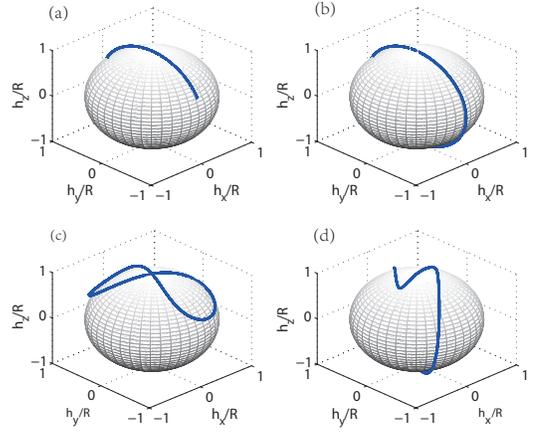}
\caption{(Color online) The paths of the Hamiltonian for 1D $2\times2$ models with $R=\sqrt{h_x^2+h_y^2+h_z^2}$. (a) and (b) show the cases of $\mathbb{Z}$-type systems, the path of the Hamiltonian and the origin are in the same plane (x-z plane in these cases). (c) and (d) show the cases of $\mathbb{Z}_2$ systems. (c) Both $h(0)$ and $h(\pi)$ are on the north pole, corresponding to $\gamma=0$ and $M=1$; (d) $h(0)$ is on the south pole while $h(\pi)$ is on the north pole, corresponding to $\gamma=\pi$ and $M=-1$.} \label{fig1}
\end{figure}

The Majorana number of the BdG classes has a correspondence to the quantized Berry phase, which is protected by the particle-hole symmetry\cite{Hatsugai}. Here we reveal this protection by examining the winding path of the Hamiltonian on the Bloch sphere. The Berry phase in the momentum space is defined as
\begin{eqnarray}
\gamma=\int_{-\pi}^{\pi}\mathrm{d}k\langle u_k|i\partial_k|u_k\rangle,
\end{eqnarray}
with $u_k$ the occupied Bloch states which are eigenstates of the Hamiltonian $h(k)$. In general, the Berry
phase $\gamma$ across the Brillouin zone is also referred as Zak phase\cite{Zak}. This phase can be expressed as $\gamma=\Omega(c)/2$, with $c$ the close loop that the Hamiltonian forms on the Bloch sphere when $k$ varies from $-\pi$ to $\pi$, and $\Omega(c)$ the solid angle of the surface enclosed by $c$. By analyzing the loop $c$ on the Bloch sphere, one can obtain the conclusion that the quantized Berry phase is protected by different symmetries for 1D $\mathbb{Z}$-type and $\mathbb{Z}_2$-type topological systems. For a 1D $2\times2$ $\mathbb{Z}$-type topological system, the existence of chiral symmetry ensures that $c$ can only be in a great circle on the Bloch sphere (Fig.\ref{fig1}(a) and (b)), hence the Berry phase can either be $\gamma=\Omega(c)/2=\pi$ or $0$. For $\mathbb{Z}_2$ systems, the Hamiltonian contains all the three Pauli's matrices, and $c$ no longer stays in a great circle (Fig.\ref{fig1}(c) and (d)). Nevertheless, the particle-hole symmetry of Eq.\ref{PHS} ensures that the Hamiltonian satisfies
\begin{eqnarray}
h_{x}(k)&=&-h_{x}(-k),\nonumber\\
h_{y}(k)&=&-h_{y}(-k),\nonumber\\
h_z(k)&=&h_z(-k). \label{hk_symmetry}
\end{eqnarray}
Dividing the loop $c$ as $c=c_1+c_2$, with $c_1$ the loop from $k=0$ to $k=\pi$, and $c_2$ the loop from $k=0$ to $k=-\pi$, one can see that these two loops are inverse to each other on x-y plane due to these conditions. As both $h_x$ and $h_y$ equal to zero when $k=0$ or $k=\pi$, $h(0)$ and $h(\pi)$ can only be on the north pole (when $h(k)>0$) or the south pole (when $h(k)<0$) of the Bloch sphere. If both $h(0)$ and $h(\pi)$ are on the same pole, $c_1$ and $c_2$ form two inverse closed loops (fig.\ref{fig2}(c)), and the Berry phase gives $\gamma=(\Omega(c_1)+\Omega(c_2))/2=0$. If $h(0)$ and $h(\pi)$ are on the different poles, $c_1$ and $c_2$ form one closed loop together (fig.\ref{fig2}(d)), and the solid angle of this loop is $2\pi$ due to the inversion symmetry between $c_1$ and $c_2$. Hence we have
\begin{eqnarray}
\gamma=\begin{cases}
0, &\mathrm{sgn}[(h_z(0)]=\mathrm{sgn}[(h_z(\pi)], \cr \pi, &\mathrm{sgn}[(h_z(0)]=-\mathrm{sgn}[(h_z(\pi)],
\end{cases}\label{BerryPhase}
\end{eqnarray}
which corresponds to the Majorana number $M=1$ and $M=-1$, respectively.

While both of the Berry phase and the Majorana number can serve as topological invariants as long as the spectrum is gapped, at the phase transition point, the gap is closed and these topological invariants become ill defined. To characterize the topological properties of phase transition points, we set
\begin{eqnarray}
k=k_0+A\sin{\theta},~~\eta=\eta_0+A\cos{\theta}\label{parameterSpace}
\end{eqnarray}
following Ref.\cite{TransitionPointPRB,TransitionPointEPL}, and define the Berry phase $\gamma_d=\int_{-\pi}^{\pi}\mathrm{d}\theta\langle u_\theta|i\partial_{\theta}|u_\theta\rangle$,
with $\eta$ the transition driving parameter. This $\gamma_d$ stands for the Berry phase along a circular trajectory around $(k_0, \eta_0)$ in the parameter space of $k$ and $\eta$, with $\theta$ the varying angle and $A$ the radius of the circular. Substitute Eqs.(\ref{parameterSpace}) to Eqs.(\ref{hk_symmetry}),  we have $h_{x}(k_0, \theta)=-h_{x}(-k_0, -\theta)$, $h_{y}(k_0, \theta)=-h_{y}(-k_0, -\theta)$ and $h_z(k_0,\theta)=h_z(-k_0,-\theta)$, which means that the integral paths from $\theta=-\pi$ to $\pi$ form two inverse closed loops for $h(k_0)$ and $h(-k_0)$, and the summation of $\gamma_d$ for $h(k_0)$ and $h(-k_0)$ always gives zero. The only exception is at $k_0=0$ or $\pi$, where $k_0=-k_0$ and the value of $\gamma_d$ is given by
\begin{eqnarray}
\gamma_d=\begin{cases}
0, &\mathrm{sgn}[(h_z(\theta=0)]=\mathrm{sgn}[(h_z(\theta=\pi)], \cr \pi, &\mathrm{sgn}[(h_z(\theta=0)]=-\mathrm{sgn}[(h_z(\theta=\pi)].
\end{cases}
\end{eqnarray}
While the change of topology of a phase transition is associated with every gap closing point in the Brillouin zone, the particle-hole symmetry ensures that these points always appear in symmetric pairs with momentum $k_0$ and $-k_0$, which gives $\gamma_d(k_0)+\gamma_d(-k_0)=0$. This leaves only $\gamma_d^+$ for $k_0=0$ and $\gamma_d^-$ for $k_0=\pi$ to determine the topology of a phase transition.

In Table.\ref{table1} we show the value of momentum $k$ and $\eta$ for different $h^{\pm}_z(\theta)$, with + (-) corresponding to $k=0$ ($\pi$). If $\eta_0$ is a topological phase transition point, $h^{\pm}_z(0)$ and $h^{\pm}_z(\pi)$ are of two different topological phases, and we can determine the Berry phase of these two phases respectively, as indicated by Eq.(\ref{BerryPhase}). Therefore $\gamma_d^+$ and $\gamma_d^-$ have different values of $0$ and $\pi$, as three of these four $h^{\pm}_z(\theta)$ have the same sign, and the other one has a opposite sign. Hence the summation of $\gamma^{\pm}_d$ is given by $\gamma_s=(\gamma_d^+ +\gamma_d^-)~\mathrm{mod} (2\pi)=\pi$ for a topological phase transition. However, if the phases of $\eta>\eta_0$ and $\eta<\eta_0$ are topologically the same, we have $\mathrm{sgn}[h^{+}_z(0)]\mathrm{sgn}[h^{-}_z(0)]=\mathrm{sgn}[h^{+}_z(\pi)]\mathrm{sgn}[h^{-}_z(\pi)]$, which gives $\gamma_s=0$. From this straightforward discussion, we can see that a $\mathbb{Z}_2$ topological phase transition can be characterized by $\gamma_s$, which is the summation of the Berry phase $\gamma_d$ around $(k_0,\eta_0)$ with a modulus of $2\pi$, where $k_0$ takes $0$ and $\pi$.
\begin{table}
\begin{tabular}{|c|c|c|c|c|}
\hline
\hline
 & $h^+(0)$ & $h^+(\pi)$ & $h^-(0)$ & $h^-(\pi)$\\
\hline
$k$ & $0$ & $0$ & $\pi$ & $\pi$ \\
\hline
$~~\eta~~$ & $~~\eta_0+A~~$ & $~~\eta_0-A~~$ & $~~\eta_0+A~~$ & $~~\eta_0-A~~$ \\
\hline
\end{tabular}
\caption{The value of momentum $k$ and $\eta$ of different $h^{\pm}_z(\theta)$}\label{table1}
\end{table}

Before moving on to the next section, we would like to point out that the situation for DIII class is not exactly the same. Although the gap closing points for such cases also come in pairs, each one of them is of a different Bloch state respectively, and they are connected by the time-reversal symmetry (TRS). Hence we also need to consider $\gamma_d$ for momentum other then $k=0$ or $\pi$ when calculating $\gamma_s$ for each state. Details will be discussed in section.IV.

\section{D Class}
Next we use our method to examine an extended version of the Kitaev model \cite{Kitaev} with additional next-nearest neighbor hopping and pairing terms, as an example of the D class. The Hamiltonian is descried by
\begin{eqnarray}
H&=&\sum_i^L t_1\hat{c}^{\dagger}_{i}\hat{c}_{i+1}+t_2\hat{c}^{\dagger}_{i}\hat{c}_{i+2}+h.c.\nonumber\\
&&+\Delta_1\hat{c}_{i}\hat{c}_{i+1}+\Delta_2\hat{c}_{i}\hat{c}_{i+2}+h.c.-2\mu\hat{c}^{\dagger}_{i}\hat{c}_{i},
\end{eqnarray}
where $\hat{c}^\dagger_i$ ($\hat{c}_i$) is the creation (annihilation) operator of fermions at the {\it i}-th site, and $L$ is the total number of sites. This model is related to the extended quantum Ising model with additional three-body interaction by Jordan-Wigner transformation \cite{NiuY,SongZ}. When $\Delta_{1,2}$ take real numbers, the Hamiltonian preserves the TRS for the spinless system, and belongs to the BDI class ($\mathbb{Z}$ type) characterized by the winding number. For the case with complex pairing terms $\Delta_1\rightarrow\Delta_1 e^{i\phi_1}$ and $\Delta_2\rightarrow\Delta_2 e^{i\phi_2}$, when $\phi_1=\phi_2=\phi$, we can eliminate the phase parameters $\phi$ by taking the gauge transformation $\hat{c}^{\dagger}_{i}\rightarrow e^{i\phi/2}\hat{c}^{\dagger}_{i}$. However, when $\phi_1\neq\phi_2$, the two phase parameters can not be eliminated at the same time, and the model falls into the D class of the tenfold way classification\cite{classification}, which is characterized by a $\mathbb{Z}_2$ number. In the following calculation, we choose $\phi_1=0$ and $\phi_2=\phi$, as we can always eliminate one of the phase parameters by choosing an appropriate gauge.
We proceed the Fourier transformation $\hat{c}_{j}=1/\sqrt{L}\sum_k e^{ikj}\hat{c}_{k}$, and the Hamiltonian then takes the form of
\begin{eqnarray}
H = \sum_k\psi^{\dagger}_k h(k) \psi_k ,
\end{eqnarray}
where $\psi^{\dagger}_k = (\hat{c}_{k},\hat{c}^{\dagger}_{-k})$ and
\begin{eqnarray}
h(k)=\left(
\begin{array}{cc}
h_z &h_x-ih_y\\
h_x+ih_y &-h_z
\end{array}\right),\label{HKitaev}
\end{eqnarray}
with $h_x=\Delta_2\sin{\phi}\sin{2k}$, $h_y=\Delta_1\sin{k}+\Delta_2\cos{\phi}\sin{2k}$ and $h_z=-t_1\cos{k}-t_2\cos{2k}+\mu$. The eigenvalues are given by
\begin{eqnarray}
E(k) = \pm |h(k)|= \pm \sqrt{h_x^2+h_y^2 + h_z^2 }.
\end{eqnarray}

\begin{figure}
\includegraphics[width=0.8\linewidth]{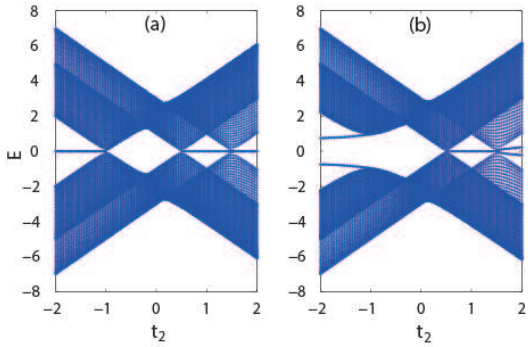}
\caption{(Color online) Energy spectrum of the D class extended Kitaev model with $\mu=1$, $\Delta_1=t_1=0.5$ and $\Delta_2=t_2$ under the open boundary condition. (a) $\phi=0$, the model belongs to the BDI class. (b) $\phi=\pi/2$, the time reversal symmetry is broken by $\phi$, and the model belongs to the D class.} \label{fig2}
\end{figure}

When $\phi=0$ or $\pi$, $h_x$ vanishes and the Hamiltonian has the chiral symmetry $\sigma_x h(k) \sigma_x=-h(k)$. After a rotation $z\rightarrow y \rightarrow x\rightarrow z$, we can calculate the winding number $\nu$ of this model, which is associated with the Berry phase as
\begin{eqnarray}
\gamma=\nu\pi~\mathrm{mod}(2\pi).
\end{eqnarray}
The winding number describes the total number of times that the Hamiltonian travels counterclockwise around the origin in the x-y plane, with the momentum $k$ varying from $-\pi$ to $\pi$. In Fig.\ref{fig2}(a) we show the energy spectrum with $\phi=0$ under open boundary condition. The number of degenerate zero modes is $4$, $2$ and $0$ in the region of $(t_2<-1\cup t_2>1.5)$, $(0.5<t_2<1.5)$ and $(-1<t_2<0.5)$, which correspond to winding number $\nu=2$, $1$ and $0$, respectively.

\begin{figure}
\includegraphics[width=0.8\linewidth]{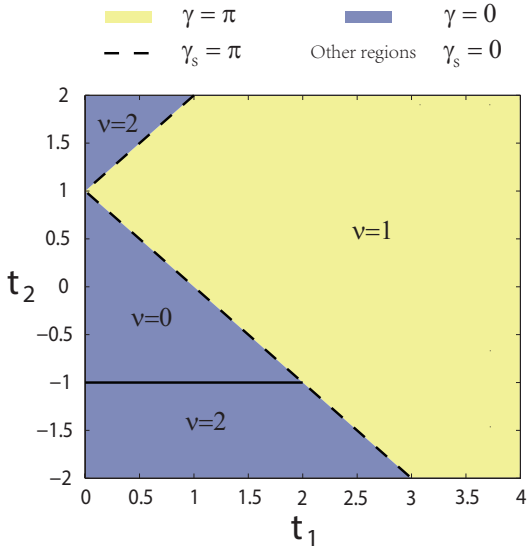}
\caption{(Color online) Phase diagram of the D class extended Kitaev model, with $\Delta_1=t_1$ and $\Delta_2=t_2$. The winding number when $\phi=0$ is labeled in the diagram. The Berry phase for model with a general $\phi$ is shown by the colour of the regions, with blue (dark) region of $\gamma=0$ and yellow (light) region of $\gamma=\pi$. The dashed and solid lines are the phase boundaries for the system with $\phi=0$, and only the phase transition marked by the dashed line survives when $\phi\neq 0$ or $\pi$. The Berry phase $\gamma_s$ is also shown in the diagram.} \label{fig3}
\end{figure}

For a general $\phi$, the existence of $h_x$ term breaks the chiral symmetry, and the topology of the system can no longer be described by the winding number $\nu$. In Fig.\ref{fig2}(b) we can see that the 4-fold degenerate zero modes in (a) split into two branches, and merge into the bulk states without a gap closing. However, the 2-fold degenerate zero modes still exist when $0.5<t_2<1.5$, and the system has two topologically different phases characterized by the Berry phase $\gamma=0$ or $\pi$. In Fig.\ref{fig3} we show the phase diagram of our model, with both winding number at $\phi=0$ and the Berry phase for general $\phi$ labeled on it. The region of $\nu=1$ when $\phi=0$ has a Berry phase $\gamma=\pi$ when $\phi\neq0$ or $\pi$, while regions of $\nu=0$ and $\nu=2$ merge together for a general $\phi$, and have $\gamma=0$.

To characterize the topological properties of the phase transition points, we calculate the Berry phase $\gamma_s$ defined in the previous section by choosing $t_2$ as the transition driving parameter, and the result is also shown in Fig.\ref{fig3}. We can see that $\gamma_s=\pi$ when the system is at $\mathbb{Z}_2$ phase transition points, and $\gamma_s=0$ for other situations.

\section{DIII class}
In this section we study the DIII class, which is the other 1D BdG class with $\mathbb{Z}_2$ topology. DIII class requires TRS with the time-reversal operator $\mathcal{T}$ satisfied $\mathcal{T}^2=-1$, so the model must have half-integer spin, and the Hamiltonian in momentum space is described by at least a $4\times4$ matrix. Nevertheless, the TRS indicates that the eigen-states always come in pairs, and both members of such a pair have the same Berry phase \cite{FuKane}, which characterizes different topological phases. According to Kramers theorem, the two eigenstates (labeled by the Kramers index $\kappa=I,~II$) in one pair have degenerate energies at time reversal invariant momenta of $k=0$ or $\pi$. In this case, the topological transition of this model is characterized by the closing of the gap between different pairs of eigenstates, and the corresponding Berry phase fulfills $\gamma^I=\gamma^{II}$, where $\gamma^{I(II)}=\int_{-\pi}^{\pi}\mathrm{d}k\langle u^{I(II)}_k|i\partial_k|u^{I(II)}_k\rangle$ is defined for each of the Bloch states $u^{I(II)}_k$ of a Kramers pair. Similar to the case of the D class, we can also define Berry phases around gap closing points to characterize a topological phase transition.
\begin{figure}
\includegraphics[width=0.8\linewidth]{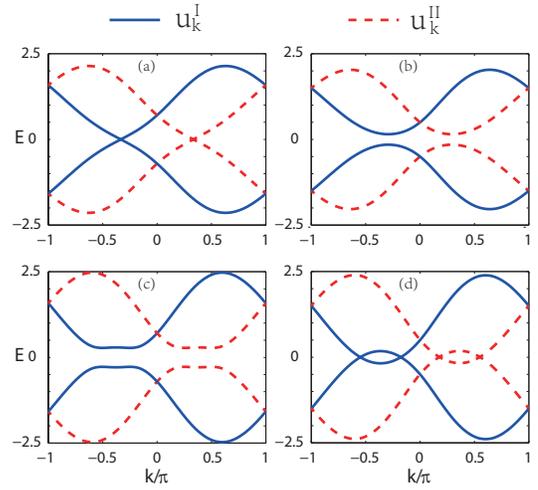}
\caption{(Color online) Energy spectrum of the DIII model with $\Delta_p=1$ and $\mu=0.5$. (a) $\Delta_s=0.5$, $\alpha_R=\sqrt{6}/3$; (b) $\Delta_s=0$, $\alpha_R=\sqrt{6}/3$; (c) $\Delta_s=0.5$, $\alpha_R=1.2$; (d) $\Delta_s=0$, $\alpha_R=1.2$. The blue solid lines and the red dashed lines are for $u_k^{I}$ and $u_k^{II}$ of a Karmers pair, respectively. The spectrum of these two Kramers pairs (the upper pair and lower pair) are symmetric about $E=0$, due to the particle-hole symmetry.} \label{fig4}
\end{figure}

As an example, here we investigate a specific DIII model with no any additional symmetry other than TRS and PHS\cite{Budich2}. This model consists of two time-reversal copies of Kitaev¡¯s p-wave chain, coupled by a Rashba spin-orbit term and augmented by an ordinary (s-wave) superconducting pairing term that
competes with the p-wave coupling. The BdG Hamiltonian of this model reads
\begin{eqnarray}
h(k)&=&(1-\mu-\cos{k})~s_0\otimes\sigma_z+\Delta_p\sin{k}~s_0\otimes\sigma_y\nonumber\\
&&+\alpha_R\sin{k}~s_x\otimes\sigma_0+\Delta_s~s_y\otimes\sigma_y,
\end{eqnarray}
here $s$ are the Pauli's matrices act on spin space, $\Delta_s,~\Delta_p$ are the SC parings, and $\alpha_R$ is the Rashba spin-orbit coupling.
This Hamiltonian fulfills the particle-hole symmetry of Eq.\ref{PHS}, and the TRS $\mathcal{T}^{-1}h(k)\mathcal{T}=h(-k)$ with $\mathcal{T}=is_y\mathcal{K}$ for spin-1/2 system. This model contains two Kramers pairs connected by the PHS, as shown by the spectrum in Fig.\ref{fig4}.
\begin{figure}
\includegraphics[width=0.8\linewidth]{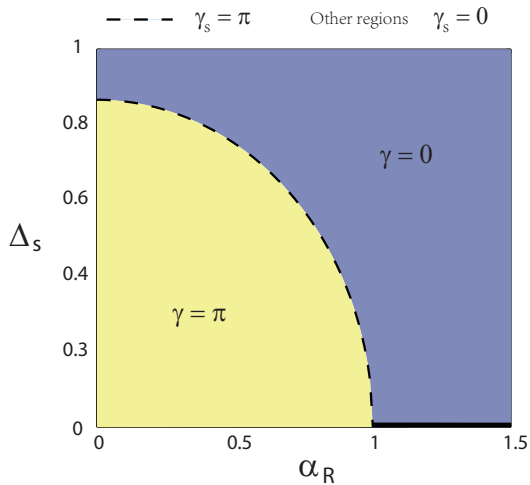}
\caption{(Color online) Phase diagram of the DIII model, with $\Delta_p=1$ and $\mu=0.5$. The blue (dark) region is of $\gamma=0$ and yellow (light) region is of $\gamma=\pi$. The dashed line is the phase transition boundary. The solid line is a metallic region, while the phases on different sides of this region have the same Berry phase.} \label{fig5}
\end{figure}

The topological properties of this model is characterized by the $\mathbb{Z}_2$ number defined in Ref.\cite{Budich2}, which is equivalent to the Berry phase $\gamma^{I(II)}$ of each Bloch state in a Kramers pair \cite{FuKane}. We calculate $\gamma=\gamma^{I(II)}$ numerically and the result is shown in Fig.\ref{fig5}. The system has two topological different phases with $\gamma=0$ or $\pi$, corresponding to the $\mathbb{Z}_2$ number $\nu=1$ or $-1$, respectively. By diagonalizing the Hamiltonian, we can determine the phase boundary (dashed line in Fig.\ref{fig5}) separating topologically different phases by the following gap closing conditions:
\begin{eqnarray}
&&1-\mu-\cos{k}=0;\nonumber\\
&&\Delta_s^2+(\alpha_R^2-\Delta_p^2)\sin^2{k}=0.\label{gapless}
\end{eqnarray}
For $\Delta_s = 0$, we find that the gap is also closed when $(1-\mu-\cos{k})^2=(\alpha_R^2-\Delta_p^2)\sin^2{k}$, as marked by the solid line in Fig.\ref{fig5}.

Next we choose $\Delta_s$ as the transition driving parameter and consider the Berry phase $\gamma_d^{I(II)}$ of gap closing points. Due to the PHS, the gap closing points always appear in pair of opposite momenta, and the two points in such a pair are of $u^{I}_k$ and $u^{II}_k$ respectively, as shown by the spectrums in Fig.\ref{fig4}(a) and (d). As the Berry phase here is defined for $u^{I}_k$ and $u^{II}_k$ separately, the Berry phase $\gamma_d$ of these pairs can not cancel out each other because they are of different Bloch states. Hence the total Berry phase of phase transition point $\gamma^{I(II)}_s$ shall take summation of $\gamma_d^{I(II)}$ of each gap closing point of $u^{I(II)}_k$. In Fig.\ref{fig5} we also show the $\gamma_s^{I(II)}$ defined in the parameter space of $k$ and $\Delta_s$, and the topological phase transition on the dashed line is characterized by $\gamma_s^{I(II)}=\pi$.
On the other hand, we have $\gamma_s^{I(II)}=0$ in the metallic region of $\Delta_s=0$ and $\alpha_R>1$, as there are two gap closing points for each member of the Kramers pair as exemplified in Fig.\ref{fig4}(d), and the summation of $\gamma_d$ of them always gives zero with a modulus of $2\pi$.

\section{summary}
In summary, we have studied the topological
properties of phase transition points of 1D $\mathbb{Z}_2$ topological systems. We examined the symmetry-protected Berry phase of the BdG classes, and
defined the Berry phase for phase transition points in
the parameter space of system¡¯s momentum and a transition
driving parameter. We studied two different extended versions of the Kitaev model of class D and DIII, and demonstrated that the topological phase transitions can be characterized by the
introduced Berry phase around the gap closing points,
which takes the value of $\pi$ for $\mathbb{Z}_2$ topological phase transition, and $0$ for other situations.
Our theory provides a way to classify topologically different
phase transitions by directly studying the properties
of the phase transition point of 1D $\mathbb{Z}_2$ topological systems.
\begin{acknowledgments}
S. C. is supported by NSFC under Grants No. 11425419, No. 11374354 and No. 11174360.
\end{acknowledgments}

\end{document}